\documentstyle[12pt,epsfig]{article}

\def \be {\begin{equation}}
\def \ee {\end{equation}}
\def \beq {\begin{eqnarray}}
\def \eeq {\end{eqnarray}}

\def \00 {$0^+\to 0^+$}

\def \SM {Standard Model}

\def \KK {Kaluza-Klein}

\def \BLG {baryo-lepto genesis}

\def  \SS {superstring}

\begin{document}

\vspace{0.1in}
\begin{flushright}
{ To the memory of \\
 E.P.Kuznetsov,  A. V. Samoilov,\\
 Yu.P. Nikitin, V. V. Ammosov}\\
\end{flushright}

\vspace{0.1in}

\begin{centering}

{\large \bf Neutrino On The Possible New  Time Structure.}\\

\vspace{0.1in}

{\bf D.S. Baranov$*$ and G.G.Volkov$**$}\\

\vspace{0.1in}

{\it $*$ Joint Institute of  High Temperatures
Moscow, Russia}\\
{\it $**$ St. Petersburg Nuclear Physics Institute
Gatchina, Russia}\\
{\it  Neutrino Light Collaboration} \\
{E-mail:baranovd@rambler.ru; ge.volkov@yandex.ru\\}


\begin{abstract}
We continue to study the problems of discovering  new temporal and spatial properties of neutrinos from the
point of the  possible multi-dimensional  extension   the D=(3+1)- special theory of relativity.  It is neutrino that can connect our Universe with new types of the matter, the new Universe.  The possible discovery with neutrino new  structure of the Time can confirm these ideas. However,  the neutrino experiments aimed at   new phenomena search can lead us to paradoxes related with
 the limits of applicability of the theory of  relativity  which demands special studies.
 This new phenomenon one can call by  "Neutrino Paradoxes in Theories of Relativity".
 As examples of such paradoxes one can   illustrate on the neutrino experiments MINOS,OPERA et al devoted to   the measurements of neutrino velocity. As the interpretation of their results are model-dependent, by our opinion, the main goal of the observation a possible new time structure in these experiments is not reached.  As the solutions to this problem it seems to us the way of holding cycle of long base neutrino experiments at high(super-high) energies which requires further debate in the literature. In addition to our discussions one can remark that  the questions of neutrino physics ($M_S  \rightarrow  O(1-20\, TeV- range)$) could be directly related to the collider physics at the LHC.
\end{abstract}

\end{centering}
\newpage
\tableofcontents
\section{Introduction.}
The neutrino speed measurement experiments are
the continuations of the classic light speed measurement experiments
have been done in range of the solar planet system(Ole Roemer-1676),
in star system (James Bradely,1728) and, at last,
on the  Earth( Lois Fizeau, 1849),....
The finite light speed measurement has led to the revolution
in  the humanity consciousness  and eventually led to a new understanding of the visible universe.
In 1998-2005, there were  a lot of excited discussions at CERN \cite{AV},\cite{AV2},\cite{Paris}
about the possibilities to perform the  neutrino experiments
to test the super-luminal neutrino hypothesis and
to find new phenomena beyond the \SM.
From one hand the idea of such experiments was associated with the hope to understand
the role of the $V-A$- weak interactions,
the  quark-lepton family symmetry,
the neutrino space-time properties and
to observe some indications on a new vacuum structure existence outside of
the Weak Scale, {\it i.e.} in the region $1/R \sim (0.1-20)TeV$.
From another hand
the general trends of this idea has been related to the possible
existence some extra space-time non-compact dimensions of the universe.
In this context it would be first serious encounter with the dual conception
between the physical phenomena of micro-cosmos and of universe.
One of the main goals is to find with neutrino some new  spatial and temporal
structures that might explain the formation of our visible $D=(3+1)$-universe
with all its space-time and internal symmetries
which could  be only a part of a vast Universe
filled with other kinds of matter.
The main difficulties of such experiments related to
the possible relativity principle paradoxes have been discussed.

\section{ Light in the Evolution of the  Time  Structure.}

Up to date there are several reasons for considering the possible extension
of the observed $D = (3 +1)$- space-time  and its symmetries.
In the first instance
a number of fundamental problems in the field of high energy physics
encountered in the Standard Model as well as in modern approaches $D = 10$ super-string theories / D-branes, 11-dimensional M-theory and 12-dimensional F-theory
have to be taken into account.

The further great advance in Physics could be related to
the   progress  of a modern  mathematics:
multidimensional Riemannian geometry, new theories of numbers,
algebras and symmetries.
Especially, we expect the powerful influence of this progress in
the understanding  of the basic Standard Model symmetries and beyond,
in   mysterious neutrino physics and its possible relation to
the  dark matter and dark energy, in high dimensional Gravity and Cosmology.
It could lead to  the further development in
the understanding of our ambient space-time,
the origin of the  Poincar$\acute{e}$-Lorentz symmetry with
the  matter-antimatter asymmetry,
the geometrical basis of the fundamental  physical characteristics:
$EM$-charge, color, spin, mass.

To date it seems naturally expect that there is necessity
to expand our knowledge about new geometrical Riemann and
tensor structures in the multi-dimensional spaces to achieve
the better understanding of the Standard Model dynamic approaches.
These new geometric objects could be associated with
some new types of external symmetries (symmetry vacuum),
which  could allow  to create
a "reasonable" (renormalized) quantum field theories
in multidimensional spaces with $D> 4$ and to construct
the multidimensional generalization of the D-dimensional pseudo-Lorentz groups,
what is an essential feature of the progress in the understanding
the principles of the general relativity theory.
The differential equations for
the propagation of waves in a hypothetical multi-dimensional space-time
could  have the third-  or higher degree with some exotic properties
as a result of observing new symmetries.

There have been presented enough experimental arguments
that the Special Theory of Relativity is being  related to
the electromagnetic charged matter
what can be applied only in $D = (3 +1) $ Minkowski space-time.
The special relativity theory was formulated on the basis of axioms comprise
the relativity principle, absolutism and the finite speed of light.
Galilean symmetry group has been extended to
the group of Lorentz transformations, and
Poincar$\acute{e}$ translation group, and
the absolutism of time transformed into absolutism of light.

Due to  light  synchronization   in stationary system  one can determine time globally. The link  time and spatial coordinates  between  two inertial systems moving relatively to each other at a constant speed is defined by  Lorentz transformations. These transformations can be built on the principle of maximal and constant speed of light and, therefore, locally determine the geometric structure of the electromagnetic vacuum, which is reflected in the fact that these  transformations leave the four-dimensional interval
\begin{equation}
ds^2=g_{\mu\nu}x^{\mu}x^{\nu}=c^2dt^2-dx_1^2-dx_2^2-dx_3^2
\end{equation}
invariant.

The progress with understanding the light speed  axiom  was gone  in the direct accordance with the progress in the study the Euclidean plane axioms where  changing
the axiom of parallel lines had led after very long period
to the discovery of Lobachevsky spaces and Riemann geometry, and
eventually had led to the discovery of the special theory of relativity
in Minkowsky $D=(3+1)$-space-time.

It worth to note that the light speed maximum axiom can be interpreted primarily
in close connection with the properties of electro-magnetic vacuum of our
visible Universe.
In Maxwell theory the absolutism of light speed is confirmed by
identification  the velocity of e.m. waves  with
the basic fundamental constants  characterizing
the internal electromagnetic vacuum structure:
\begin{equation}
c=(\mu_0 \epsilon_0)^{-1/2}
\end{equation}

The concept of the light speed absolutism in the observable Universe was
especially emphasized in the analysis of Einstein's fundamental ideas
of the special relativity theory.
The question of the new forms of the  matter existence
other than electromagnetic did not arise at those days !
The  mysteries neutrino was embedded in physics later.

Attempts of solving the problem of the Standard Model incompleteness
were converted into multi-dimensional geometry where
there could be a hypothetical sterile Matter (Dark Matter) with
his "invisible" radiation in addition to
the observed electromagnetic-charged matter
and for the description of which
there can appear the necessity to generalize some
$D = (3 +1)$ axioms of the relativity theory.
The basic idea of the such new phenomena discovering
could  be associated with the neutrino (or dark matter)
since  their unique properties could also spread in the space-time with
one or two extra dimensions, $D = (4 +1)$ or $D = (4 +2)$, respectively.
The corresponding  extension of the $D=(3+1)$- special theory of relativity can lead to
the possibility that for neutrino  the 4-interval $ds^2$ is not invariant, what could be happened if neutrino can expand in $D>4$-space-time with the extra 'topological" cycle \cite{AV}. This cycle could be related with some new spatial as well as with some temporal
neutrino properties.

To the contrast of  the spatial and temporal properties of neutrinos
with respect to the similar properties of charged quarks and charged leptons
there is a room to consider the observed three neutrino states
as a single quantum field in the space of dimension 6, that is,
with 2 additional non-compact dimensions, and, in accordance with ternary
complexity, one can imagine  three implementations of neutrinos as
a "particle" - "anti-particle" and "anti-anti-particle" (ternary neutrino model)
\cite{AV},\cite{AV2}, \cite{Paris},\cite{D6} in analogy with
the 4-dimensional Dirac electron-anti-electron(positron) theory( see \cite{Dirac},
 \cite{Majorana}, \cite{Weyl}
and for review \cite{Bog},\cite{Pal}).

\section{The EM-Charge Matter in the $D=(3+1)$ Space-Time }.

 Relativistic quantum  electrodynamics was formulated on the basis on  the  internal $U (1_{EM})$- and external Lorentz
$SO (3,1) $+ Poincar$\acute{e}$ (translation) group symmetries  of the gauge boson and fermion  fields- photon and electron/positron,  respectively.
The internal symmetry is related to the local and global conservation of
electric charge $Q_{EM}$. The external symmetries   reflect the fact that our space is isotropic and homogeneous what we observe  in the form of the law's  conservations of such fundamental parameters as the angular momentum, momentum, energy,  mass and life time at rest {\it etc} in  the $D=(3 +1)$ universe.

In theory of electron+positron  there can be some   duality links between
the  space-time geometrical structure and  fundamental    properties of the particles \cite{Paris},\cite{D6}.
For example, if one knows the fundamental properties of the particles
one can get the information such as the ambient space-time dimensions.
So, the four-dimensional $D=(3+1)$ space-time
with external Lorentz/Poincar$\acute{e}$ quantum electrodynamics symmetry
correctly corresponds to the possible quantum states - electrons+positrons -
having the following internal properties:
two spin states plus two charge conjugated states, electron/positron.

The finite  discrete group symmetries related
to the  $C$-,$P$-,$T$- transformations   make this link more subtle putting
it  finally  to the fundamental theorem of $CPT$-invariance (see for example \cite{Bog}.

The $CPT$-invariance proved in such spaces for local quantum theory
gives the very important results such as
the  equality of the particle and antiparticle masses(and life time):
\be
m(\Psi)\, =\, m({\bar{\Psi}})\,\,\rightarrow\,  binary \,\,
 CPT \,-\,invariance.
 \ee
 Similar to the role of the axiom of constant speed of light in the definition of the global time  the conservation laws of these symmetries allow us to determine such  fundamental parameters globally in the whole space-time.
 CPT-invariance allows to correctly define globally  the
concept of a particle and its antiparticle in the whole $D=(3+1)$-space-time.

In this approach the $CPT$-invariance and $Q_{em}$-conservation law can be
the prerogatives for Minkowski $R^{3,1}$- space-time
where the $SO(3,1)$  Lorentz group (Poincar$\acute {e}$) symmetry and
$U(1_{em})$ gauge symmetry\cite{AV},\cite{AV2},\cite{Paris} are  valid.
So, we want to emphasize that the proposal about  the duality
between the electric charge conservation and CPT-invariance
can be valid in our Minkowski space $D=(3 +1)$ only,
but for the hypothetical interactions of the $Q_{em}$-charged matter with the new exotic matter, these arguments  are not valid any more.

In this approximation the observation of effects with CPT invariance violation
and/or  with $Q_{em}$ charge non conservation could indicate
some new exotic geometrical vacuum structures at the smaller distances beyond
the weak interaction region  or/and
the existence of some global extra dimensions in Universe.

This observation will help us to extend the concepts of particles and antiparticles in the ternary case with 3-neutrino specie, for which a new type of hypothetical  conjugation $ C_{D=6}$ can extend the concept of anti-world to the high-dimensional analogue of $D=(3+1)$-CPT-theorem \cite{Paris}, \cite{D6}.
The observable violation of the conservation laws in $D=(3+1)$ must be associated with  some additional
geometry and tensor  structures of  vacuum  and can be linked to the appearance  some hypothetical  phenomena like new interactions, new particles,...
In Standard Model this was vividly illustrated by physics of  $K^0-\bar K^0$, $D^0-\bar D^0$, $B^0-\bar B^0$,....mixing.

The possible Majorano neutrino nature (see for review \cite{KLAP}) among
the all other kinds  \SM  Dirac charged fermions
prompts another dynamics of \BLG  based on
the composite fermion Dirac matter structure created from
more simple pra-fermions like Majorano neutrino sterile matter
filling  the extra dimensional world- Meta-Universe.

The fundamental conception such as idea is related with attempts
to figure out a common the $Q_{em}$ charge Dirac matter creation mechanism
with enough reasonable assumption that mechanism is such as
must give a duality between the $Q_{em}$ conservation and
$CPT$ invariance:
\begin{eqnarray}
CPT-\,invariance \qquad \leftrightarrow \qquad (Q_{em})\,
\,charge\,\, conservation,
\end{eqnarray}
{\it i.e.} the invariance of $CPT$ in $D=4$ space-time  means
the $Q_{em}$ charge conservation and vice versa.

Thus, if this kind duality exists, the CPT-invariance violating processes
should accompanied by the electromagnetic charge violation too.
May be,  it could be one of the reasons
why we have not saw some rare decays such as the proton decay.
In this case, the idea of grand unification symmetries without
the electromagnetic charge origin explanation
is not enough to solve  the proton stability problem.

Also a similar problem could be related with
searches for the rare flavour-changing decay channels
such as $\mu \rightarrow e + \gamma$, $\mu \rightarrow 3e$ etc .
First, one must solve the origin of the quark-lepton families problem.

In such approach one can propose a mechanism of
the geometrical electromagnetic charge $Q_{em}$ origin
and the Dirac complex  matter from more fundamental  pra-matter \cite{Paris}.
The Majorano neutrinos with $m_{Dirac}(D=3+1)=0$ could be some representatives
of a new matter (sterile or dark matter?).
The idea can be applied to the further attempts
to solve the baryon asymmetry of universe problem
linking such question with an origin
$Q_{em}$ charge matter in $D=(3+1)$-space-time.
There is one very remarkable fact
\begin{eqnarray}
|Q_p+Q_e|<10 ^{-21}
\end{eqnarray}
which can indicate the unique origin of
the proton(quarks) and electron. It suggests an existence of a hypothetical interaction into
high dimensional space connecting the Dirac-charged fermions to
the pra-matter.
This interaction could be  based on a new symmetry beyond Lie groups and
can provide the universal electron/proton  non-stability mechanism \cite{Paris}.
We add two extra dimensions to illustrate a possible mechanism of
this kind interaction with a mass scale near
$M_S \sim 10-20 TeV$- region \cite{Paris}.

Note, that the 3-color "up"- and "down"- quarks states interacting
via $SU(3^C)$-gauge color bosons at the corresponding  distances
embedded into $D=(3+1)$-space-time is connected to the problems of
a new quantum charge "color" and fractional magnitudes of electromagnetic charge
$Q=\pm 1/3,\pm 2/3$ origin explanations.
We suppose that these problems could be closely related to a  possible
extension of the electromagnetic vacuum substructure and its link to
the origin of the 3-quark-lepton families.
One could consider some extra compactified dimensions
what could change the foam structure of electromagnetic vacuum
to find a new quantum number geometrical sense due to its confinement property.
Thus,one should to produce the integer values of the charged leptons and
the fractional values of quarks by unify way
to find electromagnetic charge creation mechanism in universe.

\section{ Neutrino About  New Time Structures of Universe.}

 Exclusive properties of three neutrinos  could point out the existence
of a new vacuum, with properties different from the properties of
the electromagnetic and color vacua.
Moreover, it can give some information about the symmetry of
this hypothetical  vacuum that might be associated
with the exceptional properties of the three neutrino states-ternary
symmetry \cite{Paris},\cite{D6} in addition to the spin.
This new ternary symmetry could shed light on the SM dark symmetry:
\begin{equation}
N(Color)=N(Family)=N(dim. R^3)=3.
\end{equation}
So the three types of neutrinos can be described by
a single 6-dimensional wave function and it would imply the existence of
two additional dimensions with appearance in $D=6$ the fundamental discrete space-time symmetries.
It must be emphasized that the assumed charged matter ternary symmetry
must be broken in $D=(3+1)$-space-time  with all the attendant circumstances.

Opposite to the 3-neutrino masses one can see
the charged leptons and charged quarks grand mass hierarchies
increasing with the number of the families from one to the third:

\begin{eqnarray}
\begin{array}{ccc}
m(e)\approx0.5    MeV &  m(\mu)\approx 106  MeV  &  m(\tau) \approx  1.7  GeV  \\
m(up)\approx3.5   MeV &  m(c)  \approx 1250 MeV  &  m(T)    \approx  175  Gev \\
m(down)\approx5.5 Mev &  m(s)  \approx 150  MeV  &  m(B)    \approx  4.5  GeV  \\
\end{array}
\end{eqnarray}

Also one can see the reverse hierarchy of life times of
the electromagnetic charge particles according to
increasing the number of the generations:
$1\rightarrow 2\rightarrow 3\rightarrow  ...$.
There are some peculiarities related to the  mass limits  what could be important for our interpretation of
the  weak interaction region as a boundary between two  vacua: electromagnetic  and new hypothetical in $D=6$-space-time.

The first peculiarity requires to postulate
the minimal possible mass in EM-vacuum: electron mass ?
Then the "up"- and "down"- quark masses could be expressed through
the electron mass and the number of colors :
\begin{equation}
1/2(m(up)+m(down)) \approx N_c^2 m(e)
\end{equation}.
Then the next peculiarity is related to a trend of upper limits on the masses
with increasing number of the generation.

Under this circumstance it will be important to clarify
the following problem: does the fourth generation exist or doesn't?
Some superstring models possessing the hypothetical family symmetry expect
the fourth quark generation having some exceptional properties\cite{MNV3},\cite{MNV4}
In this case one could consider the quaternary extension of ternary hidden symmetry:
\begin{equation}
N(Q-3Color +L-1Color )=N(Fam.-3 + Ex.Fam.-1)=N(Dim. R^{3,1})=4.
\end{equation}
The experimental observation of the fourth quark generation could support
the idea about real role of weak interactions in the \SM and in universe:
"screening" at the  very small distances beyond the weak interaction
region $r \leq 10^{-17} cm$.
Thus one can suggest that the electromagnetic vacuum could be defined
by the light speed and by the minimal Dirac mass
possible for the stable electromagnetic Universe.

 One of the our space-time extension possibilities could be
due to a new "topological cycle"($\tau,\xi$) existence and it
could be described by  independent component such as
new "time" coordinate ($\tau$)\cite{AV},\cite{AV2}.
In this scenario, the question what is the real time raises again.
These ideas implementation should require the construction of
the Universe new geometric representations and, in particular,
to find the Riemann metric tensor and might be
another geometrical and tensor structure invariants of extended space-time.
In fact, the hypothesis of the second "time coordinate"
might be considered as a convenient way to describe
the possible extension of the neutrino spread laws different from
those projected by special relativity,
for example, the light speed maximality principle.

One of the main difficulty of
the study the neutrino intrinsic and space-time properties connected with
the considerable discrepancy between the huge experimental data
for the processes with neutrinos as products of hadron's decays and
very small amount of the processes where the space-time properties of
neutrinos clearly manifested .
If the analysis of the myriad of the neutrino channel meson decays
restore the energy and angular spectrum of the neutrino collapse
the further motion evolution of this collapse can contain significant
uncertainty  ( see for example series of the articles devoted to the study
for the formation of neutrino beams in accelerators \cite{BAR},\cite{BAR2},\cite{SAM},\cite{AFON},\cite{K2K},\cite{NuMI}, \cite{SPS},\cite{SPS2}).
Further only a tiny fraction of neutrinos in these collapses observed
in neutrino channels can be identified via the interaction of neutrinos
with detector. The ratio of the accelerator produced  neutrinos in the collapses
to those could be observed in the neutrino detector can be,
depending on the experimental conditions, the order of $\sim 10^{7-10}$.

This is especially important to review the samples of
long base-line  neutrino experiments in K2K-SuperKamiokande   \cite{SKAM},
FNAL NuMi-MINOS \cite{MINOS} and
SPS CNGS-LNGS OPERA \cite{SPS},\cite{OPERA}, \cite{BRUN}.
The CNGS beam is obtained by accelerating protons to 400 GeV/c and ejecting
ones into neutrino channel as two spills, each lasting $10.5 \mu s$  and
separated by $50ms$.The SPS CNGS cycle is 6 s long.
Each spill contains from 2100 bunches  with the time substructure
$3+2=5 nsec$ and  intensity $\sim 10^{10}POT$.
The resulting neutrino collapse is formed at the neutrino channel along
a distance of $\sim 1000 m $\cite{SPS}, \cite{SPS2}.
The total statistics used for analysis  reported in this paper \cite{OPERA}
was $\sim 15 000$ events (from $\sim 60 000$ total events) detected in rock
and in detector, corresponding to about $10^{20}$- protons on target collected
during the 2009, 2010, 2011 CNGS runs  and the estimation of
the  total work-time is about $5 \times {10}^7$ sec.
So the total number of spills could be about  $< \sim 10^6$ and each spill
produces $< \sim 0.01 \nu $-event in detector or in rock ( the exact
numbers one can see in \cite{BRUN}).
It gives very complex problem to restore the total information about
the all parameters of neutrino collapse spectrum.
Naturally the question Whether is there a chance to synchronize
neutrino events in such experiments to within less than time of extraction
{\it i.e.}  $\leq 10.5 \mu sec$ ? \cite{AV},\cite{AV2}.
Sufficiently, Do we know well the spatial and temporal properties of neutrinos
to achieve such  synchronization accuracy?

This occasion can bring the any kind paradoxes caused by
incorrect experimentalist understanding
of the space-time behaviour dynamics of the neutrino collapses based
only on extremely small recorded statistics of the detected neutrino events.
The main paradox of such experiments is that the results of long term studies
become to be equivalent to the following inference:
what  had been assumed that it was received.
The opportunity of a wrong interpretation of the ambiguous experimental results
 makes the modern experimental neutrino physics is very
complex and raises such experiments at the level of art.

It is well known that neutrino experiments consist of three phases:
the neutrino collapse production process, its space-time spread
through the matter and the possible interaction of the collapses
with the detector material.
The neutrino collapse dynamics moving in space-time is another
major challenge because of the proposal that the neutrino is
an another kind of matter representative significantly different
from the electromagnetic matter.
This suggests that neutrinos could spread in accordance with
the new vacuum structure kinematics.
Despite the existence of three quark-lepton generations
the three states forming a single wave field of $D=6$-space-time evolution
might be assumed and would be
described by the corresponding wave equation.
In a ternary model the neutrino wave field could have
the own charge - "neutrino light" \cite{Paris},\cite{D6}.
 In this approach the neutrino field could be distributed according to
the motion equations different from the equations used
in $D=(3+1)$- geometry defined by the Lorentz group symmetry.
It can give some new additional interpretations of the processes related to
the well known neutrino oscillations.

 The possible extra dimensional geometry existence can lead to the circumstance
that the neutrino waves could spread by geodesic lines different from
the geodesic lines of the  electromagnetic charged particles
(see for example, \cite{Koka}).
Appears from the above the neutrino flow cannot conserve
in the $D=(3+1)$-space-time.
It could be a reason of disappearance of neutrino flows at a distance.

In the article \cite{AV} some neutrino experiments were proposed
to observe the possible our space-time expansion comprising another cycle
characterized by its fundamental speed which could be much faster than
electromagnetic light.
The last assumption was supported by some arguments to solve
the horizon problem in cosmological models \cite{Mof}.
Neutrinos due to their outstanding properties available in both cycles and
the electromagnetic light speed maximality principle does not work any more.
In particular, the new multi-dimensional geometrical spaces have
the projective symmetries the understanding of them could help us to visualize new universes.
Another factor is that the space-time expansion can carry out
the introduction of new topological cosmic cycles.
It means that these topological cosmic cycles may have own fundamental
parameters such as  "speed",  "mass", "charge",...

Therefore,
to check the hypothesis that neutrinos spread different than the light
the experiment based on the possibility of measuring the neutrino speed
depends on the parameters
that might be related to the fundamental another cycle properties
has been suggested, and
we expected that dependence on such parameter one could get the neutrino speed:
$v_{\nu}>c_{light}$ ( see the interesting discussions
in \cite{MA},\cite{MA2},\cite{NA}, \cite{GIA}, \cite{Wolf}).

Such experiments could prove the existence the new vacuum and extra dimensions
directly but this way  involves a very delicate element
associated with synchronization "almost invisible" neutrino.

In the classical experiments to measure the new fundamental constant
the validity limits of the special relativity theory need to be understood.
For our approach it was necessary to examine on what setting might have changed
the neutrino velocity value if it really has a link to the new vacuum.
The latter implies that there should be in minds some method of the possible
$D=(3+1)$ space-time expansions with the corresponding metric tensor forms
for such  models.
Otherwise such experiments can lead to the logical paradoxes.
In fact, such experiments can meet the challenge of measuring the absolute
velocity or absolute motion or something else.

\section{Non-Compact Large Extra Dimensions and Emergence of EM-Charge.}

The main experience we have got
from \KK, \SS/D- branes, $D=11M$, $D=12F$- theories   and
from the study of the Riemann and tensor structures of
the high dimensional Cartan symmetric and Berger-non-symmetric spaces  is
the compact small  and the non-compact  large dimensions are
closely connected to the origin of internal and external space-time symmetries,
respectively, in corresponding theories (see discussion in \cite{VOL3}, \cite{VOL4}).

The compact small dimensions are connected with
the origin of internal symmetries.
The role of the compact Calabi-Yau spaces was perfectly illustrated in
the 5-superstring dual theories.
Correspondingly,  non compact  large dimensions are related to
the extra  space-time symmetries of the our ambient world.
For the \SM \, this  circumstance could be very important since we suppose that the problem of three neutrino species could be solved by adding the some global
non-compact dimensions to our $D=3+1$ space-time\cite{AV},\cite{Paris}.
So the  family symmetry appearance can be related to
the large non-compact extra dimensions like
it was happened with two "families", particle-antiparticle, and
was proved by Dirac  relativistic equations
for the $D=3+1$- Minkowski space-time.

In the past a lot of publications has been devoted to the possibility to solve
the three family problems through
the internal gauge family symmetries introduction.
Let note that in super-string approach
the $N=1SUSY$ $SU(3)_H\times U(1)_H$ gauge family symmetry appears
with $3+1$ quark-lepton family \cite{MNV3},\cite{MNV4}.
The possible fourth family must have the exceptional properties
since this family is singlet under $SU(3)_H$-symmetry in this approach .
This broken family gauge symmetry could be responsible for
the mechanism of $CP$-violation in $K-$,$D-$,$B$- meson decays \cite{MNV}, \cite{MNV2}.
Thus one can see the common  grand problem of
the flavour mass hierarchy of quarks and charged leptons, family mixing,
$CP$-violation
that cannot be solved without understanding the role of
the $(V-A)$-weak interactions.

There is also the very important difference between
the three charged quarks/leptons and three neutrino states:
Dirac-Majorano space-time nature \cite{Dirac},\cite{Majorana}, \cite{Weyl},
their masses  and etc.
We can expect that for Majorano neutrino species
the global family symmetry is exact...
To explain the ambient geometry of our world with some extra
infinite dimensions one can consider the our visible world
(Universe) as just a subspace of the Meta-Universe with the pra-matter having
new quantum numbers different from already known in our world.

The visibility of a  new world  phenomena
is determined by the our understanding of the SM structure   and its
consequences for the Cosmology processes.
The Majorano neutrino can travel in this Universe!\cite{AV}
To make it available we should introduce a new space
time-symmetry with the usual D-Lorentz symmetry generalization.
In this case the region  $ \leq M_S \sim (10-20) Tev$ could
be considered as a "boundary" of a new world \cite{Paris}.

This proposal suggests an existence  in high dimensional space
a class of prs-fermions connected with our fermions through a new interaction
which is based on a new symmetry beyond Lie.
Such interaction could predict a non-stability of the electrons or protons.
We can illustrate   a  scale of such an interaction,  taking for example,
two extra dimensions. To make estimations we are in the situation in which
occured E.Fermi.
So we can follow to his ideas which he used for construction
the four-fermion weak theory with coupling constant $G_F$, which dimension is
$[G_F]=[M^{-2}]$.
So we can start from multi-fermion $D=6$ Fermi Lagrangian
the corresponding Fermi constant $G_{F_S}$,
which should have a the dimension\cite{Paris}:
\begin{equation}
[G_{F_S}]=[M^{-4}]
\end{equation}

In our opinion this coupling constant dimension corresponds
to a new  interaction that propagator could have a form
like $1/[P(q,M_S)]$, where $P(q)$ could be a polynomial of $4$-th degree.
Such as propagator form corresponds to a new  $D=6$-metric tensor.
So, for the tree level calculations of the quark or charged lepton decays into
 pra-fermions $\nu_S$
\begin{eqnarray}
q \mapsto   n \, \nu_S, \qquad e^{\pm} \mapsto  n \, \nu_S
\end{eqnarray}
can get the following estimation for the partial width
\begin{eqnarray}
\Gamma(e/q \rightarrow n \nu_S) \sim O(g_S) \cdot\frac{m_{e/q}^9}{M_S^8},
\end{eqnarray}
where   $m_{e/q}$ is the electron mass, and $M_S$ is a new  mass scale related to
the hypothetical interaction in extra dimensional world what could be associated to the some new symmetries.
What is very interesting that we can construct the universal mechanism of
the decays for the all known quarks -$u,d,s,c,b,t$- and
charged leptons- electron, muon, $\tau$lepton- into the $EM$-invisible matter.
To get the lower boundary for $M_S$ let's compare the partial width for electron decay with the life time of muon in frame of $D_4$-Fermi interactions:

\begin{eqnarray}
\frac{\Gamma(e \rightarrow 3 \nu_s)}{\Gamma(\mu \rightarrow e\nu\bar \nu)}
=O(g_S/g)\frac{m_e^9}{m_{\mu}^5}\frac{M_W^4}{M_S^8}
 \nonumber\\
\end{eqnarray}
From the  lower boundary on the  electron life time one can get the following
upper boundary for $M_S$:
\begin{equation}
M_S \geq O(g_S/g)\cdot (10-20) \cdot M_W.
\end{equation}
This boundary has the universal magnitude what one can check from
searching the baryon violation processes of neutron
 \begin{equation}
 N \mapsto 3 \nu \qquad  or \qquad
  N \mapsto \,n  \nu_S .
 \end{equation}

Apart from charge violation decays we can expect also
the the $CPT$-invariance violation processes.
For example, the $M_S$ magnitude estimation can be get from the
$K^0$-$ \bar K^0$- mass difference:

\begin{eqnarray}
\delta_m=|m-\bar m| \sim  \frac{m^5}{M_S^4} <  10^{-15} GeV.
\end{eqnarray}
This estimation show that the $M_S $ can be also in $1-10 TeV$ region.

\section{The Paradoxes of  Theory of Relativity in Long Base Neutrino  Experiments.}

The measurements of neutrino speed on  the accelerator experiments
can be associated with  Fizeau experiment to measure the speed of Light.
Opposite to Fizeau ideology ($\{2\times 8.5  \times 720 \times 24.5\} km/sec$)  in the  neutrino  projects  \cite{AV}  there must be studied three  main discrepancies  what are related to the some  experimental and theoretical  ambiguities.
In contrast to the Fizeau experiment to measure the speed of neutrinos was a very daunting task to identify and synchronize the departure of neutrinos or neutrino wave collapse formed during the release of the accelerator proton bunch on the target. The second discrepancy  was related to the understanding the structure of neutrino collapse formed in the neutrino channel and getting all its parameters. The third important discrepancy was  linked to the driving  dynamics of neutrino fronts propagating over long distances, {\it i.e.} to represent  its  evolution during the flight from  accelerator till detector .  To solve the  third problem one should  have the information about possible some new spatial - temporal properties of neutrinos,{\it i.e.}  that is  to construct or have some sorts of models explaining the physical reasons why  the neutrinos properties    could be beyond some principles of special theory of relativity, in particularly to overcome  the  speed of light. If you do not accept this ideology  the experiments of measurements the neutrino velocity will involve with attempts to measure the "absolute motion"( Aristotle,  Galileo Galilei).

The main conclusion from this discussion is how to measure correctly speed of the objects with the properties completely different from the electromagnetic media (new time structure, synchronization). To make such experiments one can coincide themselves to the paradoxes of measurement the absolute movement. In electromagnetic media- universe and vacuum- only light can have the property of absolutism, $c$ is invariant fundamental constant . Between energy and wave length there exists the quantum link:
$E_{\gamma}\cdot \lambda_{\gamma}=c \cdot h$, where  $h$ is Plank constant.  If neutrino is not related to the supersymmetry \cite{DVV} or some unknown yet now phenomena
in the frames of the special theory of relativity there appear the ambiguous situation. Absolutism of neutrinos?  In special theory we know how to synchronize the events in electromagnetic media.

Now there can appear grand question how to synchronize the events in a new vacuum media. In this regard, in our projects, \cite{AV} we were looking for those neutrino  observable parameters that could link the neutrino with a new vacuum in which the laws of the $D=(3+1)$- electromagnetic universe could not  be valid any more.
 Other vacuums different than  electromagnetic can  have  new fundamental properties what we suggested to check in the experiments  related to the measurements the propagator properties of neutrino like its velocity.

 The extra dimensional (non trivial) generalization of the Lorentz group could  imply the existence of another boost and possible extensions the concept of the time, even its  structure. To find the confirmation of existing of exotic  vacuum structure  beyond the electromagnetic one should look for the experimental measuring parameters  of neutrinos  what could connected to the other hypothetical world. Or we can meet to the Aristotle - Galilei absolute movement problem. We know enough well the  ways to synchronize  electromagnetic events on the long distances  using  the light. But the many neutrino phenomena we don't know yet well to be sure that we can synchronize correctly the neutrino events. In  neutrino experiments  made during some last years  there was realized only the first part of the  projects \cite{AV} what was based on conclusions  of the the first measurement the neutrino velocity in 1977 FNAL\cite{KAL} and in 1987
 $[{\bf 24}]+[{\bf 5]=[{\bf 11}KII+{\bf 8}IMB+{\bf 5}B]}+[{\bf 5}M]$)-neutrino events  getting from SN87A and detected in KAMIOKANDAII, IMB, BAXAN, Mont-Blanc \cite{SN87A}.  From that observations there has been done the main conclusion  that  the magnitude of the speed of neutrino should be very closed to the corresponding magnitude of   the light speed.

Since neutrino could link both worlds for neutrino the  principle of absolutism of light is not valid more and  we
can take  the  energy as possible such parameter. The other possible parameter could be related to the sources of neutrinos since it gives the information about  the region of neutrino production on the smaller and smaller distances.

 In this choice it will be important to study the possible spatial and temporal  structures of neutrino fluxes having the wide energy spectrum,   producing from certain sources and moving on the different distances.
We can compare  the ideology of  neutrino measurement velocity experiments  \cite{KAL}, \cite{SN87A},\cite{MINOS},\cite{OPERA}, \cite{ICARUS}  to the our conceptions \cite{AV},\cite{AV2}, \cite{Paris}.
As examples we can consider  the productions of neutrino fluxes in SPS CERN from regular extractions of proton bunches in which  the energy of proton beams equal to  $E_p=400GeV$ with regular extractions during $3nsec$ with separation $500 nsec$ and intensity about and $10^{11}POT $. For each extraction one can estimate the  corresponding  neutrino
energy spectrum of $10^{8,9}$ neutrinos with the primary  period about $3 nsec$. For us it will be important to make the analysis  of time and spatial expanding of this bunch on some different distances:
$83 km,366km,732 km,1464 km$. In line with the hypothesis about the existence of $(1+1)$- extra dimensions we suggested that the  expanding of neutrino fluxes  depends
on some parameters, $\{ t,L,E_{\nu},k_i(r),c_i(y)\}$, where $t,L,E_{\nu}$ are ordinary parameters what we can measure in the  experiments, parameters $k_i$ are related to the type of sources of neutrino (muon, $\pi$- and  $K$- mesons, $charm$, beauty, - quarks ,..., parameters $c_{i}$ should be directly linked to the fundamental characteristics of a new vacuum,  depending on the type of a new hypothetical  metric tensor. As a result of this approach the spread of the neutrino collapses will be completely different from the flow of electromagnetically charged particles, {\i.e.}  the expansions of neutrinos must be beyond the laws determined by standard Lorentz $D=(3+1)$-metrics\cite{AV}.
As we approached the speed of the neutrino is not absolute characteristic for the electro-magnetic vacuum there was made the assumption  that the neutrino energy could one of the  parameters binding our vacuum with a new vacuum. We can propose  that the speed of neutrino could effectively depend on the neutrino energy what we were searching through  ternary or quaternary extensions of $D=(3+1)$-metrics what can be constructed in $R^n$-spaces,
($n \geq 5$)  \cite{D6}, \cite{VOL}:
\begin{equation}
ds^2=f_1(y_a)ds_{3,1}^2+f_2(x_{\mu})ds_y^2.
\end{equation}

  As a possible variant, one can consider that the speed of neutrino
is the product of electromagnetic charged particles could have some deviations from the speed of light:
$ (v_{\nu}/c-1)_{i} \approx k_i \cdot (E_{nu}/M_S)^2$, where parameter $k_i$ is determining by   the region on neutrino production, for example, in our examples  we  consider two cases, neutrino from $\pi$ - and $K$-meson decay for which $r\sim 1/m_{\pi}$ and $r\sim 1/m_K$, correspondingly. In our scenario, the behaviour of the neutrino velocity at super-high energies could be in accordance with the formula
$v_{\nu}\rightarrow c_{new} [E^2/(E^2+M_S^2)]$, where new hypothetical velocity constant could be much more than light speed, $c_{new}>>c $ ($c_{new}/c\geq 10^{7-8}$?).
The proposed dependence of  neutrino speed  from the energy leads to the substantial change of the spatial and temporal picture of the neutrino collapses.
 Thus according to articles \cite{AV} we consider two variants of taking parameters what could help to  observe the effects of existing a new space-time vacuum structure:
 \begin{itemize}
 \item{1. energy of neutrinos}, \\
 \item{2. parents of neutrino production}.\\
 \end{itemize}
For illustrations in the first case we consider the neutrinos formed only from $\pi$-meson decays in  CNGS neutrino channel  formed as a result of the discharge of protons on target with the energy of 400 $GeV$
 in certain  time intervals(see fig.{(1),fig.{2},fig.{3},fig.{4}} and fig.{(5)}. For this case the coefficients $k_{\pi}$ is related to the magnitude of the wave function distribution of $up$- and $down$- quarks inside the $\pi$-meson structure region.
For the second case we considered the  possible parents of neutrino-  kaons(see fig.{6}.,fig.{7},fig.{8},fig.{9} In this case we take into account both variants-energy and origin of production .  The distribution of the  neutrino energy formed from lepton- and semi-lepton- decays K-meson decays for which we take the following coefficients for  $ k_{K}/k_{\pi}\sim (m_{K}/m_{\pi})^2$ . According to \cite{AV}   the neutrino velocity effect should more significant for neutrinos produced from the heavy quark decays. This one can see also on the figures fig.{(10)} where the temporal spread of the $\nu_{K}$- neutrino collapses can be much more than it was in $\nu_{\pi}$- cases.
Combining both cases one can consider for neutrino velocity   the more strongest energy dependence like as $v_{\nu}/c \sim (E_{\nu}/M_S)^4$ ( see the fig.{11} and fig.{12}).
For illustration we can give the distributions of $\nu$-fluxes from $SN87A$ at $M_S=0.1-,0.2-,0.5-,1 TeV$(see fig{(9)}). This cosmic experiment may be the first time
gives us a hint about what the neutrinos cross  the huge spatial and time intervals  according to other laws. To draw concrete conclusions from this experiment is difficult to do. Any predictions depend on the theory of stellar evolution, and the structure of the tremendous medium through which the neutrino waves swept generated in the depths of supernova \cite{SN87A}.

\begin{figure}
\includegraphics[width=120mm]{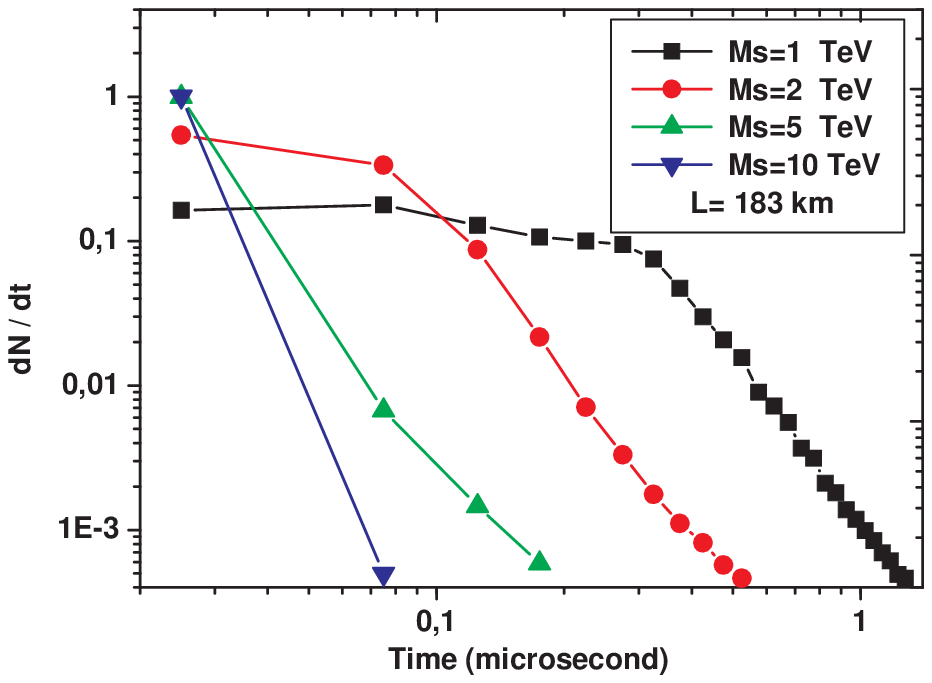}
\caption{The temporal distributions  of the intensity of the neutrino fluxes for the only one bunch
of the output proton beam at energies $E=400 GeV$ \cite{SPS}. The  duration of one bunch is equal to 3 nanoseconds, the gap between the neighboring bunches equals  500 nanoseconds. Consider the case of formation of the neutrino fluxes from
 $\pi$-meson decays. Four  distributions  $\nu_{\pi}$- fluxes at the different scales:
$M_S=1-,2-,5-,10- TeV$ on the distance $183km$.}
\label{overflow}
\end{figure}

\begin{figure}
\includegraphics[width=120mm]{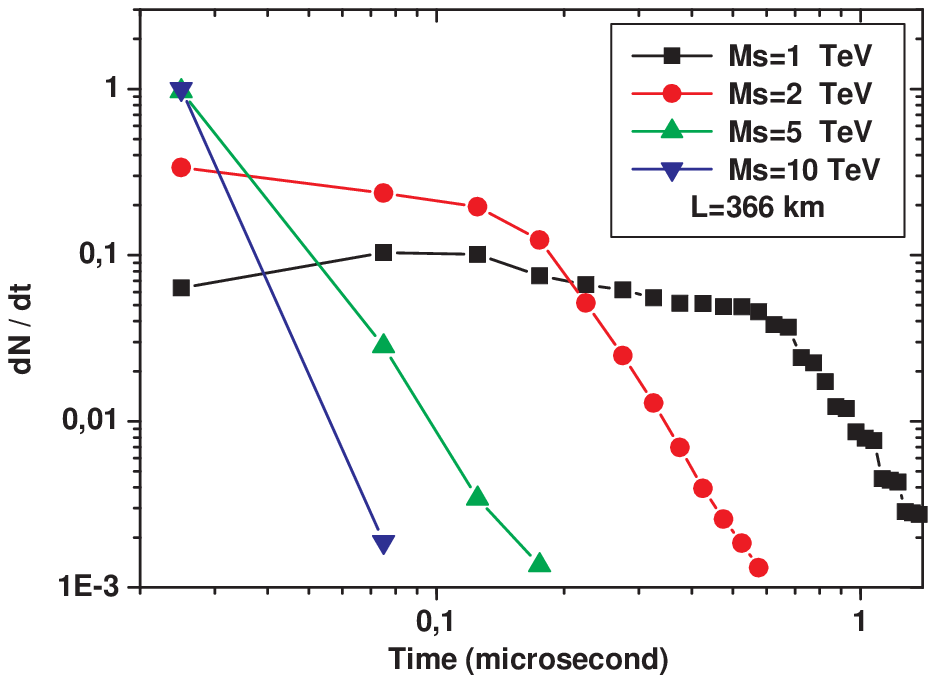}
\caption{The temporal distributions  of the intensity of the neutrino fluxes for the only one bunch
of the output proton beam at energies $E=400 GeV$ \cite{SPS}. The  duration of one bunch is equal to 3 nanoseconds, the gap between the neighboring bunches equals  500 nanoseconds. Consider the case of formation of the neutrino fluxes from
 $\pi$-meson decays. Four  distributions for $\nu_{\pi}$- fluxes at the different scales:
$M_S=1-,2-,5-,10- TeV$ on the distance  $366km$.}
\label{overflow}
\end{figure}

\begin{figure}
\includegraphics[width=120mm]{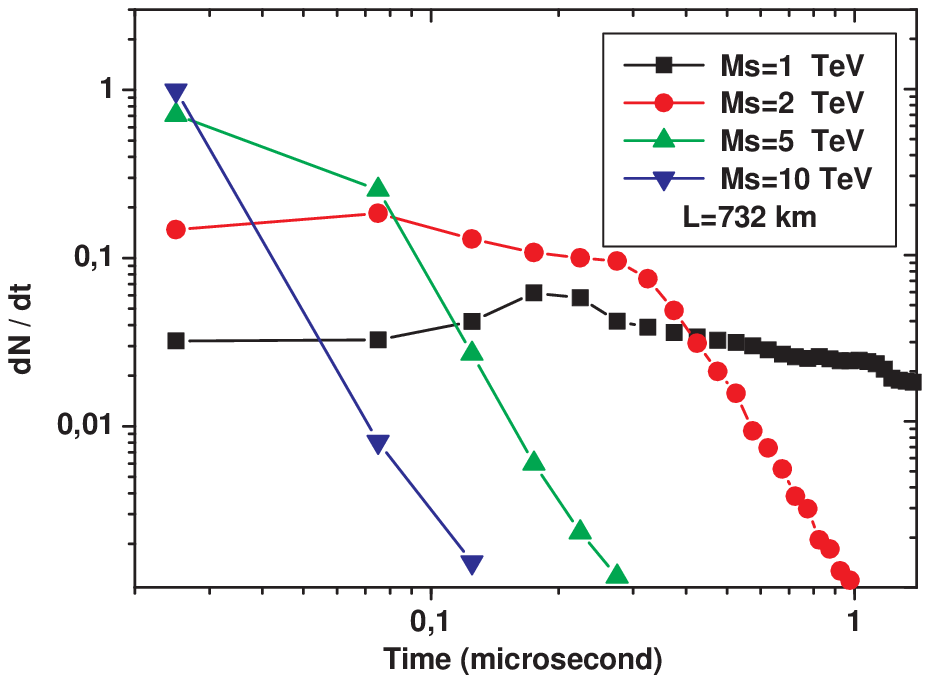}
\caption{The temporal distributions  of the intensity of the neutrino fluxes for the only one bunch
of the output proton beam at energies $E=400 GeV$ \cite{SPS}. The  duration of one bunch is equal to 3 nanoseconds, the gap between the neighboring bunches equals  500 nanoseconds. Consider the case of formation of the neutrino fluxes from $\pi$-meson decays. Four  distributions for $\nu_{\pi}$- fluxes at the different scales:
$M_S=1-,2-,5-,10- TeV$ on the distance$732km$.}
\label{overflow}
\end{figure}

\begin{figure}
\includegraphics[width=120mm]{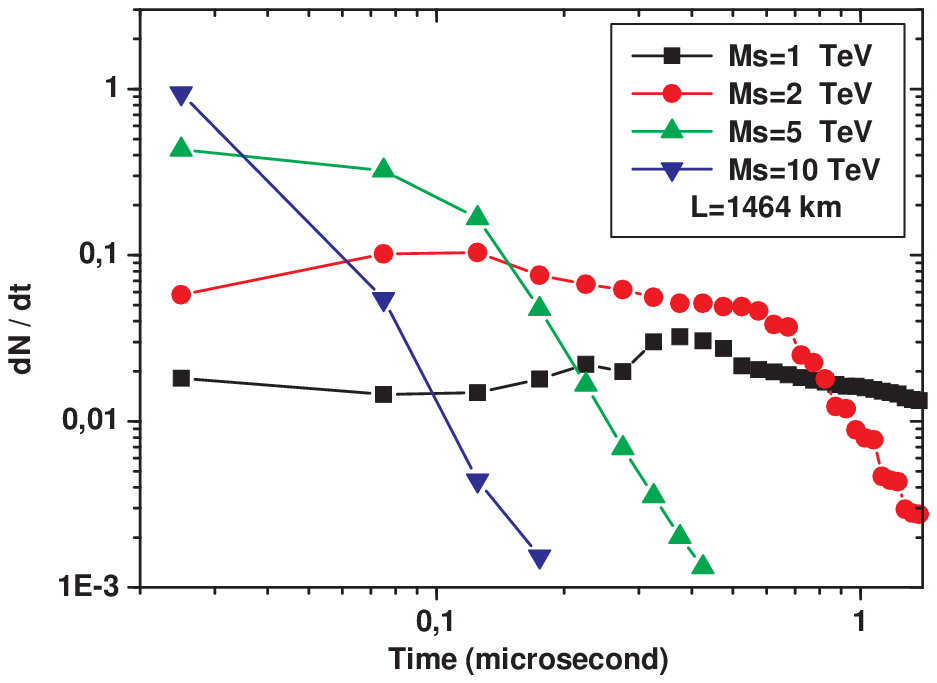}
\caption{The temporal distributions  of the intensity of the neutrino fluxes for the only one bunch
of the output proton beam at energies $E=400 GeV$ \cite{SPS}. The  duration of one bunch is equal to 3 nanoseconds, the gap between the neighboring bunches equals  500 nanoseconds. Consider the case of formation of the neutrino fluxes from $\pi$-meson decays. Four  distributions for $\nu_{\pi}$- fluxes at the different scales:
$M_S=1-,2-,5-,10- TeV$ on the distance $1464km$.}
\label{overflow}
\end{figure}

\begin{figure}
\includegraphics[width=120mm]{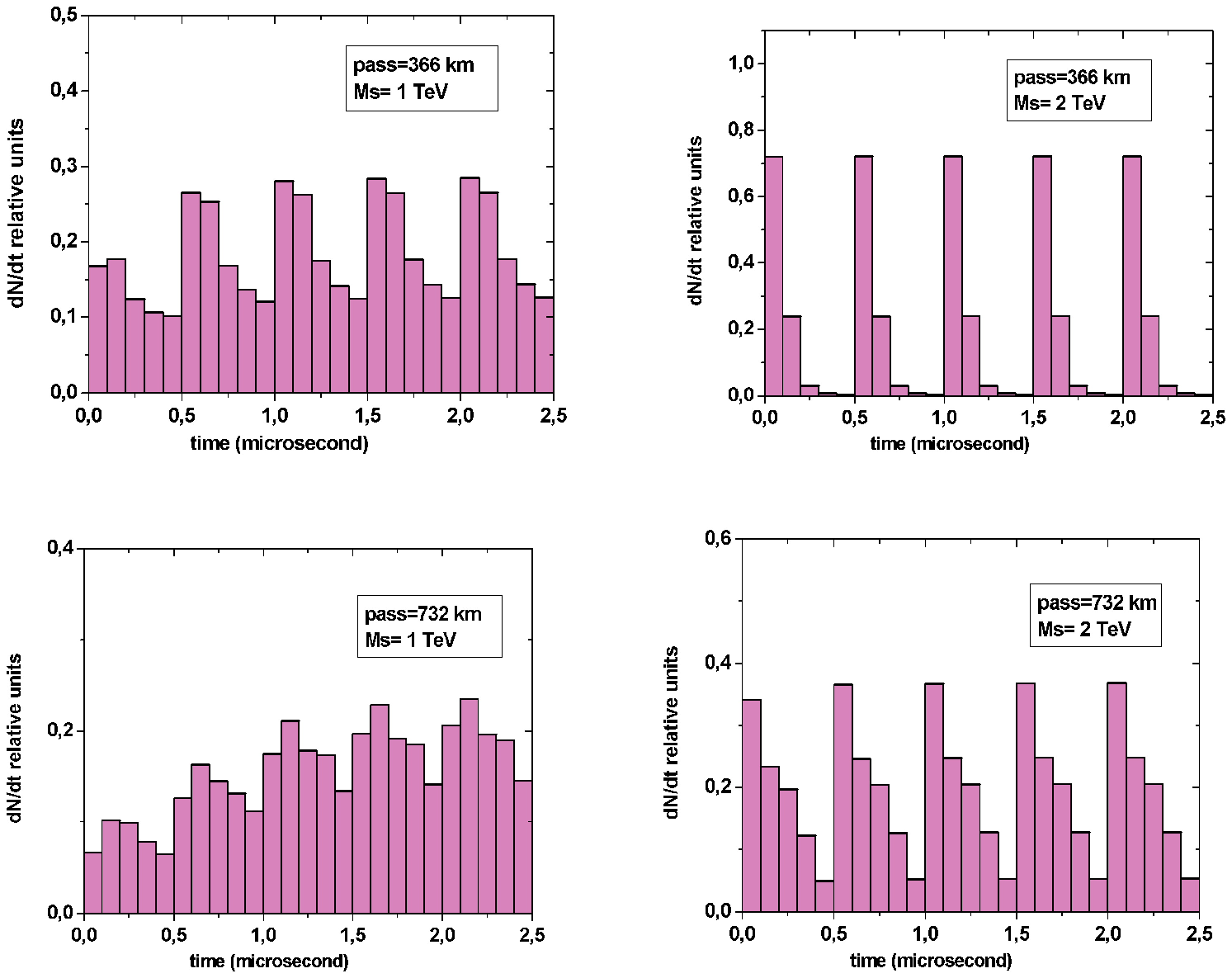}
\caption{The temporal distributions  of the intensity of the neutrino fluxes for  5-bunches from the output proton beam  with energy $E=400 GeV$ at a distances of 366km and 732 km. Ms= 1 Tev, 2 TeV.The  duration of one bunch is equal to  3 nanoseconds, the gap between the neighboring bunches equals  500 nanoseconds. Consider the case of formation   neutrino fluxes from
 $\pi$-meson decays. }
\label{overflow}
\end{figure}

\begin{figure}
\includegraphics[width=120mm]{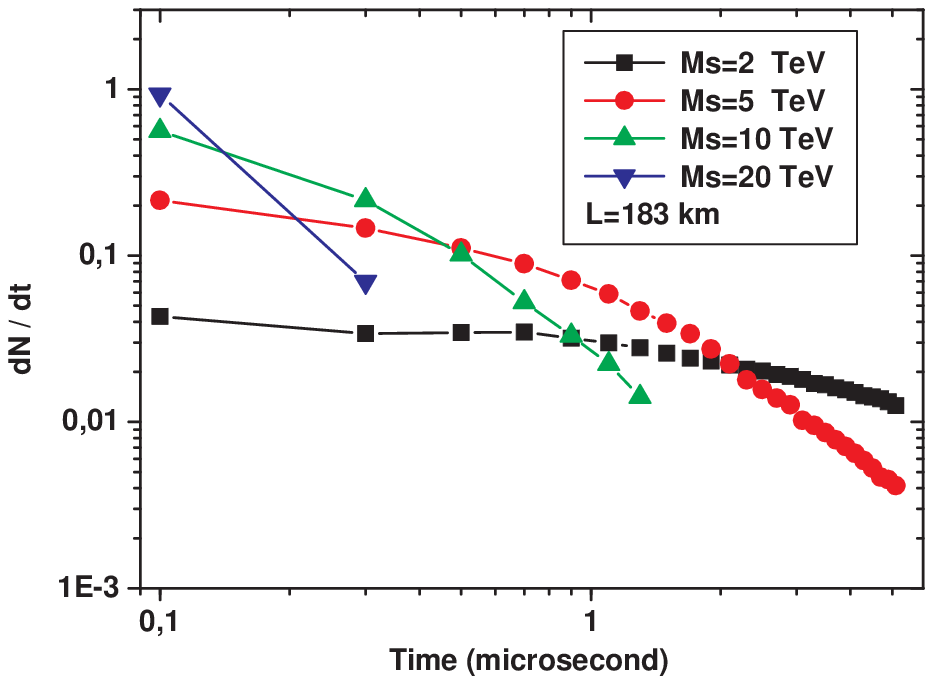}
\caption{The temporal distributions  of the intensity of the neutrino fluxes for the only one bunch
of the output proton beam at energies $E=400 GeV$ \cite{SPS}. The  duration of one bunch is equal to 3 nanoseconds, the gap between the neighboring bunches equals  500 nanoseconds. Consider the case of formation of the neutrino fluxes from
 $K$-meson decays. Four  distributions for $\nu_{K}$- fluxes at the different scales:
$M_S=2-,5-,10-,20- TeV$ on the distance $183km$.}
\label{overflow}
\end{figure}

\begin{figure}
\includegraphics[width=120mm]{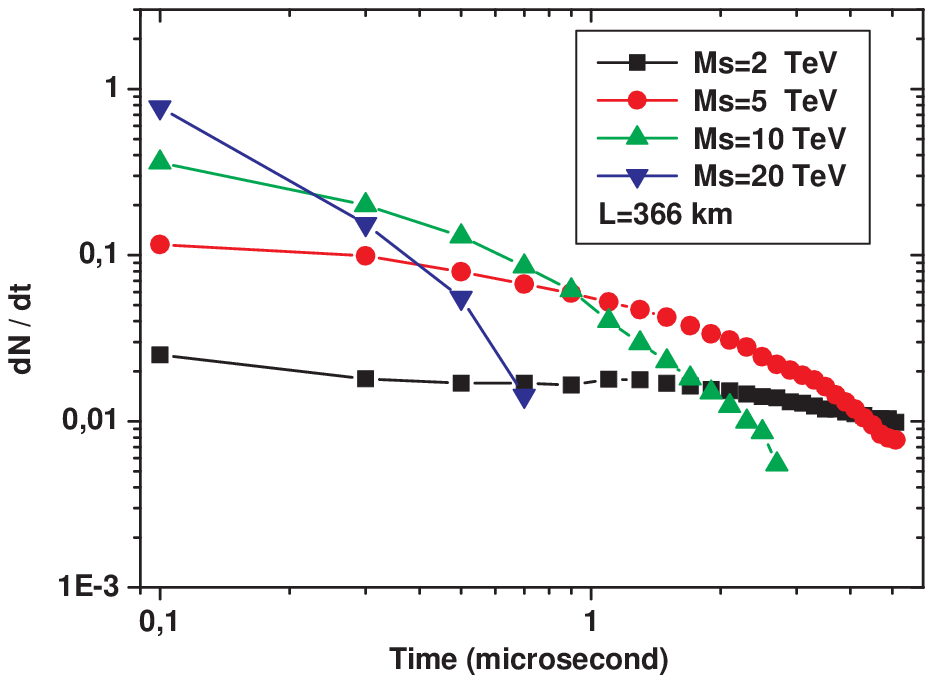}
\caption{The temporal distributions  of the intensity of the neutrino fluxes for the only one bunch
of the output proton beam at energies $E=400 GeV$ \cite{SPS}. The  duration of one bunch is equal to 3 nanoseconds, the gap between the neighboring bunches equals  500 nanoseconds. Consider the case of formation of the neutrino fluxes from
 $K$-meson decays. Four  distributions for $\nu_{K}$- fluxes at the different scales:
$M_S=2-,5-,10-,20- TeV$ on the distance $366km$.}
\label{overflow}
\end{figure}

\begin{figure}
\includegraphics[width=120mm]{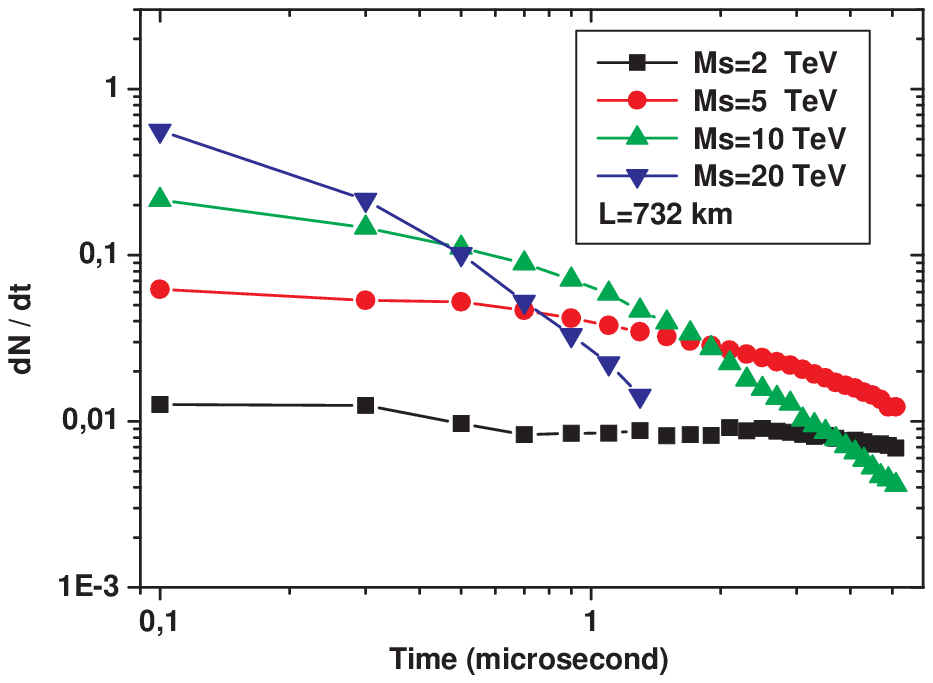}
\caption{The temporal distributions  of the intensity of the neutrino fluxes for the only one bunch
of the output proton beam at energies $E=400 GeV$ \cite{SPS}. The  duration of one bunch is equal to 3 nanoseconds, the gap between the neighboring bunches equals  500 nanoseconds. Consider the case of formation of the neutrino fluxes from
 $K$-meson decays. Four  distributions for $\nu_{K}$- fluxes at the different scales:
$M_S=2-,5-,10-,20- TeV$ on the distance $732km$.}
\label{overflow}
\end{figure}

\begin{figure}
\includegraphics[width=120mm]{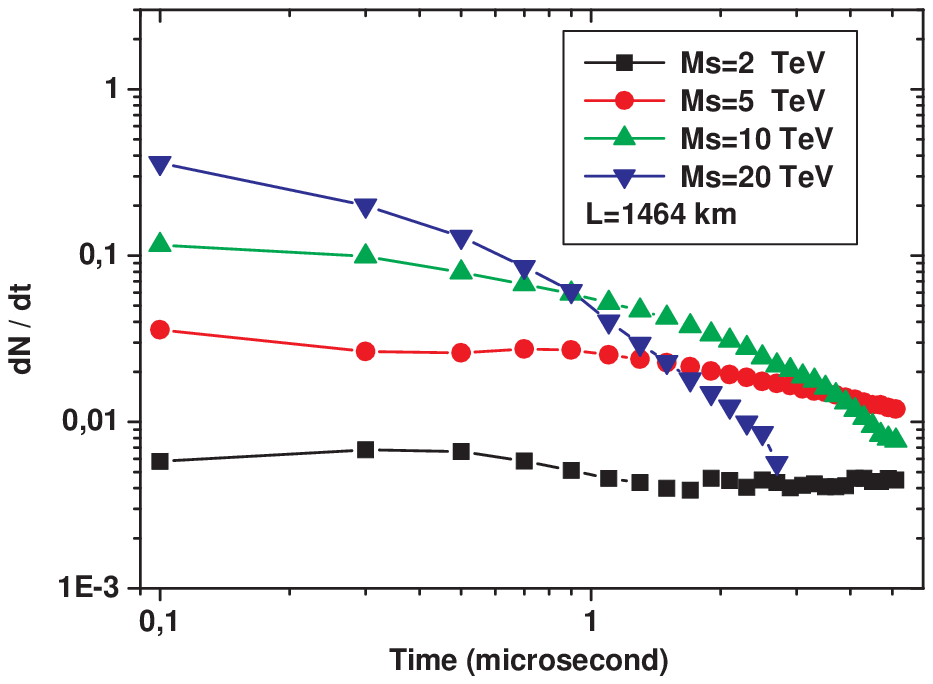}
\caption{The temporal distributions  of the intensity of the neutrino fluxes for the only one bunch
of the output proton beam at energies $E=400 GeV$ \cite{SPS}. The  duration of one bunch is equal to 3 nanoseconds, the gap between the neighboring bunches equals  500 nanoseconds. Consider the case of formation of the neutrino fluxes from
 $K$-meson decays. Four  distributions for $\nu_{K}$- fluxes at the different scales:
$M_S=2-,5-,10-,20- TeV$ on the distance $1464km$.}
\label{overflow}
\end{figure}

\begin{figure}
\centering
\includegraphics[width=150mm]{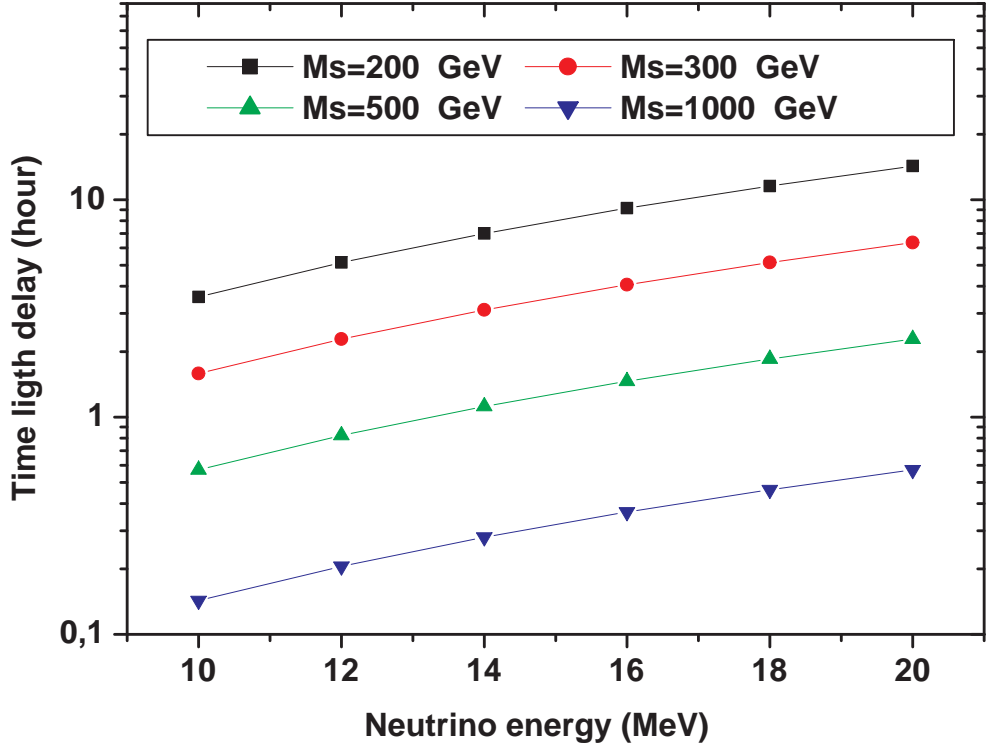}
\caption{Distributions of $\nu$-fluxes from $SN87A$ at $M_S=0.2-,0.3-,05-,1 TeV$}
\label{overflow}
\end{figure}

\begin{figure}
\includegraphics[width=120mm]{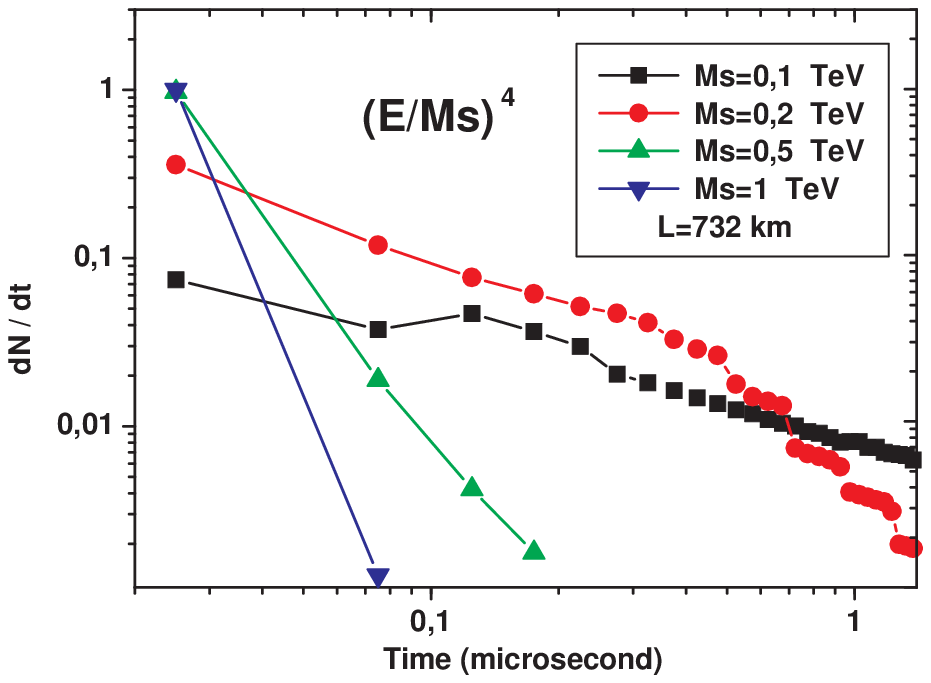}
\caption{The temporal distributions  of the intensity of the neutrino fluxes for the only one bunch
of the output proton beam at energies $E=400 GeV$ \cite{SPS}. The  duration of one bunch is equal to 3 nanoseconds, the gap between the neighboring bunches equals  500 nanoseconds. Consider the case of formation of the neutrino fluxes from
 $\pi$-meson decays. Four  distributions for $\nu_{\pi}$- fluxes at the different scales:
$M_S=0.1-,0.2-,0.5-,1- TeV$ on the distance $732km$.}
\label{overflow}
\end{figure}

\begin{figure}
\includegraphics[width=120mm]{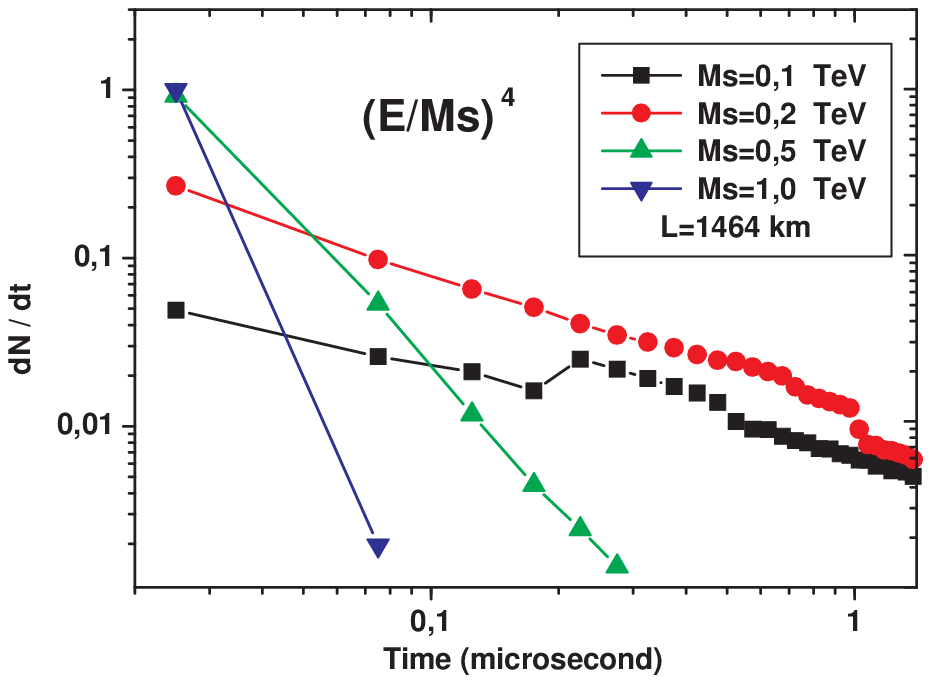}
\caption{The temporal distributions  of the intensity of the neutrino fluxes for the only one bunch
of the output proton beam at energies $E=400 GeV$ \cite{SPS}. The  duration of one bunch is equal to 3 nanoseconds, the gap between the neighboring bunches equals  500 nanoseconds. Consider the case of formation of the neutrino fluxes from
 $\pi$-meson decays. Four  distributions for $\nu_{\pi}$- fluxes at the different scales:
$M_S=0.1-,0.2-0.5-,1- TeV$ on the distance $1464km$.}
\label{overflow}
\end{figure}

\newpage

\section{Conclusions.  Towards the Super-High Energy Neutrino Experiments}

 Comparing with our ideas  in \cite{AV} the  conclusions obtained from
the data analysis of experiments \cite{MINOS}, \cite{OPERA}, \cite{ICARUS} demand a comprehensive reconsideration.
From these experiments it is not clear what they measured- neutrino or light velocity? The conclusions depend on some proposals of neutrino spatial and temporal properties.
In such kind of experiments the main accent should been  done on
the attempts of the synchronization problem solution for two "neutrino" events: production and detection.
However, to solve this problem in these experiments
it was assumed that the spatial-temporal behaviour of neutrinos is
determined by the Lorentz-Poincar${\acute e}$ group symmetry  $\psi_{\nu}\sim (\exp{i k \cdot x}$).
It wasn't taken into account
that neutrinos can have extraordinary spatial and temporal properties
different from the charged matter properties.
For example, in  ternary or in quaternary description
the following plane wave generalization can be taken in the following form \cite{D6},\cite{VOL}, \cite{VOL2}:

\begin{equation}
 \psi_{\nu}\sim \exp\{{\bf I_1} (\hat k \cdot \hat x)
  + {\bf I_2} (\hat p \cdot \hat y)+{\bf I_3}({\hat m}(p,k)\cdot {\hat s}(x,y))\}.
 \end{equation}

Therefore, the reached conclusions on the neutrino velocity measurement
cann't be interpreted unambiguously and
it was illustrated in the previous section.
It can be taken into account that these experiments collide with
the absolute movement paradoxes.
The results of such kind experiments correspond to common supposition
that the neutrino collapses propagate through the space similar to
the light fluxes.

The good time resolution was achieved in these experiments
to measure the speed of neutrinos
but the energy resolution was not sufficiently precise for
the proposed synchronization problem solution ideas.
The energy has been measured with poor accuracy at least 20 percent.
This is due to the several reasons:
\begin{itemize}
\item{1.The bad identification of the charged and neutral currents.
In the neutral current neutrino takes a lot of energy}\\
\item{2. The carried away energy of neutrinos  is  poorly defined.}\\
\item{3. The $\pi$-meson and the proton are often not distinguished.}\\
\item{4. The observation of several events, secondary events are mixed up.
}\\
\end{itemize}
the detector containing passive elements can increase
the number of problems with the neutrino energy definition.
There is also the separation problem
of the rock neutrino events.
It would be important to study the dependence of registered neutrino fluxes on
the energy and distance for the correct neutrino spectra normalization.

In fact, the experimental limits on the speed of neutrinos
obtained in these experiments can not be proven as
the maintained synchronization depends on the specific  assumptions.
In our opinion,
the problem that was posed in the works \cite{AV} requires
a whole series of serious studies under different conditions:
the wide energy range of the proton spills till some TeV,
the short and the long distances,
different time length of proton spills,
different neutrino specie, etc.

There might be very important the questions connected to
the decrease of the flux intensity  dependent on the distance
what can be occur due existence of extra dimensions and
it is naturally expected in the ternary model neutrino.
So the neutrino flux evolution with the distance and the energy
can also test the hypothesis of non-compact extra dimensions.

The considered scenarios showed that
the solution of the  neutrino phenomena mysteries
still needs new and the more detail experiments.
The our examples show that it is better to carry out
the neutrino measurements at the more higher energies.
The proton/electron stability provides
the upper bound of $M_S\sim 20 TeV$ \cite{Paris}.
Since the other models can be developed the investigation
of this class of neutrino experiments needs to be continued.

In our projects \cite{AV}, the suggestions on measurement
of some global neutrino flux motion characteristics
in spatial-temporal picture have been done.
We knew that in the case of neutrinos
the successful implementation of these projects is determined by
the our knowledge of or the correct ideas on the actual neutrino properties
as a particle and a wave or some its generalization.

Since it was suggested that
the effects of advancing the speed
depends on the  neutrino energy is natural to consider
the neutrino programs at the high or the extremely high energies.
The experiments at $\sim O(100-1000)GeV$ \cite{AV} could be already done
in the nearest future at the accelerator modern complexes of FNAL and of CERN.
One can say that such kind of experiments are linked to
the  another class of neutrino experiments at high energies opposite to
the ideology of the experiments going to measure the neutrino oscillations.
We should note that this class of neutrino experiments at super-high energy and on different bases including very long distances
can be very important for the future of the SM-physics and beyond.
For example, the experiments on the total cross section measurement
of neutrino interactions in dependence on the distance
between the source of neutrino beams and detectors might be done

 The neutrino experiments at the high or the super-high energies will be able
to expand the progress possibilities in discovering the secrets of neutrinos
significantly if they will n't not play a decisive role.
In our opinion, the two programs of
the low-energy neutrino physics associated with the study of
the oscillations,\cite{SKAM},\cite{MACRO},\cite{SOUDAN},\cite{MINOS2}, {\it etc} and
the neutrino physics at the ultra-high energies \cite{BAR2},\cite{AV}
will only complement each other.

\section*{Acknowledgements}
We  would like to express his acknowledgements to colleagues in TH-PNPI
Ya. Azimov, G. Danilov, A. Erikalov, L. Lipatov, S. Trojan and
others for very useful discussions on this subject.
Also we thank  F. Anselmo, N.  Budanov, M.Chernecov, A. Kisselev,  A.  Liparteliani,  E. Slobodyuk  for  many interesting  discussions and support.
We thank D.I. Patalakha for reading English version of manuscript and
making some useful comments.
This is the extension and more precise version of our talk on the SHQCD12-Conference, Gatchina, St-Peterburg,12-16 July 2012.

\end{document}